\documentstyle[aaspp4]{article}
\begin{document}

\title{Declining Rotation Curve and Brown Dwarf MACHOs}
\author{Mareki Honma\altaffilmark{1}}
\affil{Institute of Astronomy, University of Tokyo, Mitaka, 181, Japan}
\authoremail{honma@milano.mtk.nao.ac.jp}

\and

\author{Yukitoshi Kan-ya}
\affil{National Astronomical Observatory, Mitaka, Tokyo, 181, Japan}
\altaffiltext{1}{Research Fellow of the Japan Society for the Promotion of Science}
\slugcomment{submitted to the Astrophysical Journal Letters}

\begin{abstract}

If the Galactic rotation speed at the Solar circle is $\sim 200$ km s$^{-1}$ or smaller, which is supported by several recent studies, the rotation curve of the Galaxy could be declining in the outermost region.
Motivated by this, we investigate the effect of such declining rotation curve on the estimate of the MACHO mass and the fractional contribution of the MACHOs to the Galactic dark halo.
Using Hernquist and Plummer halo models instead of the standard halo model, we find that the MACHO mass could be significantly smaller than that for the standard halo case.
In particular, there exists a certain set of halo parameters for which the MACHO mass is 0.1$M_\odot$ or less and at the same time the MACHO contribution to the total mass of the halo is almost 100 \%.
This result indicates that a halo which consists solely of brown dwarfs can be consistent with both of the observed microlensing properties and the constraints from the rotation curve, provided the outer rotation curve is indeed declining.

\end{abstract}

\section{Introduction}
The observations of gravitational microlensing towards the Large
Magellanic Cloud (LMC) have revealed that a significant fraction of the
Milky Way's halo consists of solar or sub-solar-mass compact objects, which are called massive compact halo objects (hereafter MACHOs; Alcock {\it et al.}  1993, 1996, 1997, Aubourg {\it et al.}  1993). 
So far at least six microlensing events have been detected toward the LMC, and recent analyses have shown that the MACHO mass is about $0.5M_\odot$ (Alcock {\it et al.} 1997).
The fractional contribution of the MACHOs to the Galactic halo mass is also estimated to be $f\simeq 40\%$(Alcock {\it et al.} 1997), but may vary between 15 and 100\% (e.g., Kan-ya, Nishi \& Nakamura 1996).

The mass estimate of $\sim 0.5M_\odot$ is consistent with red dwarf stars and old white dwarfs as baryonic candidates for MACHOs, but both halos have serious difficulties.
A red dwarf halo cannot be consistent with the observations since red dwarfs do not appear in the star counts of Galactic halo fields (Graff \& Freese 1996).
If MACHOs are old white dwarfs, the initial mass function (IMF) of the progenitor stars must be strongly peaked around $\sim 2M_\odot$ (Adams \& Laughlin 1997, Chabrier, Segretain \& M\'era 1996), being completely different from the present day IMF (e.g. Scalo 1986).
Moreover, the halos of distant galaxies must be brightened by progenitor stars of the present day MACHO white dwarfs, which contradicts with the number count of distant galaxies (Charlot \& Silk 1995).
The overproduction of helium, carbon and nitrogen is also a major problem for a white dwarf halo (Charlot \& Silk 1995; Gibson \& Mould 1997).
Furthermore, the age of the white dwarfs which are cool and faint enough
to be consistent with the negative result of HDF star count (Flynn,
Gould \& Bahcall 1996) is very old (Chabrier, Segretain \& M\'era 1996).
The age estimated by Chabrier {\it et al.} (1996) is not consistent with
recent observations of the Hubble constant.

On the other hand, brown dwarfs are possible and attractive candidates for MACHOs, since it is free from problems concerning metals and star counts.
The largest problem of a brown dwarf halo is the consistency with the mass estimate of $\sim 0.5 M_\odot$.
The mass estimate is, however, based on the {`}standard{'} halo model, in which the rotation curve is flat out to the LMC.
Hence, if the standard halo model does not represent the Galactic halo well, the MACHO mass would be changed considerably.
Recently, Honma \& Sofue (1996; 1997a) revealed that the outer rotation curve of the Galaxy could be declining if the Galactic constant $\Theta_0$, which is the rotation velocity of the Galaxy at the Solar circle, is 200 km s$^{-1}$ or less, and several studies indeed suggested $\Theta$ of $\sim 180$ km s$^{-1}$ (e.g., Kuijken \& Tremaine 1994; Olling \& Merrifield 1998).
Furthermore, many observations have revealed declining rotation curves in the outer region of spiral galaxies (e.g., Casertano \& van Gorkom 1991; Olling 1996; Bland-Hawthorn et al 1997; see also Honma \& Sofue 1997b and references therein), indicating that declining rotation curves are not uncommon.

For these reasons, it is important to re-analyze the MACHO mass with {`}non-standard{'} halo models for which the outer rotation curve decline.
In this Letter we study the MACHO mass and the halo MACHO fraction based on such halo models and investigate the possibility of a brown dwarf halo.

\section{Halo Models}

Since this investigation is motivated by declining rotation curves, we focus on the halo models whose density profiles decrease faster than the standard halo model, $\rho\propto r^{-2}$, of which Hernquist model and Plummer model are common examples.
The density of Hernquist models is proportional to $r^{-4}$ at large radii, and it is known to trace the luminosity profile of elliptical galaxies moderately well.
The density of Plummer models decreases faster than Hernquist model at large radii, proportional to $r^{-5}$.
The Plummer model is known to trace the light distribution of globular clusters to some degree.
Conveniently enough, the potential-density pairs and the distribution functions are obtained analytically for both models.
Below we summarize the analytic form of the potential-density pairs and distribution functions.

\subsection{Hernquist Model}

The relative potential and the density for Hernquist model (Hernquist 1990) are written as,
\begin{equation}
\Psi_{\rm H} = \frac{GM}{r+a},
\end{equation}
and
\begin{equation}
\rho_{\rm H} = \frac{M a}{2\pi r (r + a)^3},
\end{equation}
where $G$ is the gravitational constant, $M$ is the total mass of the halo, and $a$ is the scale length.
Hernquist (1990) obtained the distribution function for this model as 
\begin{equation}
F_{\rm H} = \frac{M}{8\sqrt{2} \pi^3 a^3 v_g^3}\frac{1}{(1-q^2)^{5/2}} 
\left[3 \sin^{-1}q + q(1-q^2)^{1/2}(1-2q^2)(8q^4-8q^2-3)\right],
\end{equation}
where $q=({a \cal E}/GM)^{1/2}$ and $v_g=(GM/a)^{1/2}$.
Note that the relative energy ${\cal E}$ is defined as ${\cal E}\equiv\Psi-v^2/2$.

\subsection{Plummer Model}

The potential-density pair for the Plummer model (see Binney \& Tremaine 1987) are
\begin{equation}
\Psi_{\rm P} = \frac{GM}{(r^2+a^2)^{1/2}},
\end{equation}
and
\begin{equation}
\rho_{\rm P} = \left(\frac{3 M a^2}{4\pi}\right) (r^2 + a^2)^{-5/2}.
\end{equation}
The distribution function for this model is written as 
\begin{equation}
F_{\rm P} = \frac{24\sqrt{2} a^2}{7 \pi^3 G^5 M^4}\; {\cal E}^{7/2}.
\end{equation}

Note that both of Hernquist and Plummer models contain only two parameters, the total mass $M$ and the scale length $a$.

\section{Simple Estimate of $\tau$ and $m$}

Based on the models described above, we calculate the optical depth $\tau$ and the MACHO mass $m$.
Paczynski (1986) defined the optical depth as
\begin{equation}
\tau =\int \pi R_E^2\; \frac{\rho}{m}\; dD,
\end{equation}
where $R_{\rm E}$ is the Einstein ring radius, $m$ is the MACHO mass, and $D$ is the distance along the line of sight.
We assume that the distance to the LMC is 50 kpc, and also that all MACHOs have a unique mass $m$.
$\tau$ is related to the halo MACHO fraction, but independent of the MACHO mass because $m$ cancels out in eq.(7).
A quantity related to $m$ is the differential event rate $d\Gamma/d\hat{t}$ (Griest 1991; see also Alcock {\it et al.}1996), which may be written as
\begin{equation}
\frac{d\Gamma}{d\hat{t}}= \frac{32\pi u_{\rm T}}{m} \int\int \frac{R_{\rm E}^4}{\hat{t}^4}\; F\; dv_r dD.
\end{equation}
Here $u_{\rm T}$ is the maximum impact parameter, $\hat{t} =2R_{\rm E}/v_{\perp}$ is the Einstein ring crossing time, and $v_r$ and $v_{\perp}$ are the radial and tangential velocities of MACHOs, respectively.
We set $u_{\rm T}=0.661$ following Alcock {\it et al.}(1997).
The total event rate $\Gamma$ can be obtained by integrating eq.(8) with respect to $\hat{t}$, namely, $\Gamma = \int\; (d\Gamma/d\hat{t})\; d\hat{t}$.
The expectation of $\hat{t}$ can be obtained as
\begin{equation}
<\hat{t}> = \frac{1}{\Gamma} \int \hat{t}\; \frac{d\Gamma}{d\hat{t}}\; d\hat{t}.
\end{equation}
Note that $<\hat{t}>$ is proportional to $\sqrt{m}$ (Griest 1991), and hence, the MACHO mass can be estimated by comparing $<\hat{t}>$ with the observed one, $<\hat{t}>_{\rm obs}$.
To be conservative, we adopted $\tau_{\rm obs}=2.1_{-0.7}^{+1.1}\times 10^{-7}$ and $<\hat{t}>_{\rm obs}$ of 80 days from six events so far detected (Alcock {\it et al.} 1997).
Note that this $<\hat{t}>_{\rm obs}$ is the simple arithmetic mean of the blending-corrected timescales for the six events, and the observational efficiency is not included here (see next section for more elaborate analyses).

Figures 1 show the contours for $\tau$ and $m$ for Hernquist models and Plummer models in the parameter space of $a$ and $M$.
The distance to the Galactic center $R_0$ is assumed to be 7.5 kpc.
If the IAU standard $R_0$ of 8.5 kpc is assumed, the optical depth becomes slightly smaller, but the results are not significantly changed.
Figures 1 demonstrate that the MACHO mass can be $0.1M_\odot$ or less depending on the halo parameters.
In particular, the region for $m<0.1 M_\odot$ and the region for $\tau\sim 2.1_{-0.7}^{+1.1}\times 10^{-7}$ overlap (shadowed region in figures 1).
This fact implies that the MACHO mass can be less than $0.1M_\odot$ while keeping the halo MACHO fraction close to $\sim 100$\%.
Therefore, a halo which consists solely of brown dwarfs ($M\le 0.08 M_\odot$) is possible as long as these {`}non-standard{'} halo models are accepted.

Figure 2 shows the rotation curves modeled with the non-standard halos plus a disk and a bulge.
An infinitely-thin exponential disk is assumed for the disk component, and Plummer model is assumed for the bulge.
The scale lengths are fixed to be 3 kpc for the disk, and 0.4 kpc for the bulge, respectively.
The two rotation curves in figure 2 (full lines) correspond to the Galaxy's models with Hernquist and Plummer halo.
For the Galaxy's model with Hernquist halo, we set $M=2.0\times 10^{11}M_\odot$ and $a=13$ kpc for the halo, and the disk mass of $3.5\times 10^{10}M_\odot$ and the bulge mass of $0.8\times 10^{10}M_\odot$ (hereafter model H).
For the Galaxy's model with Plummer halo, we set $M=1.1\times 10^{11}M_\odot$ and $a=12$ kpc for the halo, and the disk mass of $4.0\times 10^{10}M_\odot$ and the bulge mass of $1.0\times 10^{10}M_\odot$ (hereafter model P).
The halo parameters for models H and P lie in the shadowed region in figure 1.

In figure 2 the observed rotation curve taken from Honma \& Sofue (1997a) is also plotted ($\Theta_0=180$ km s$^{-1}$ assumed).
The model rotation curves in figure 2 reproduce well the general tendency of observed rotation curves, except for a local deviation in the vicinity of the Sun.
The local density of the exponential disks assumed here are 51 and 58 $M_\odot$ pc$^{-2}$, being in agreement with the studies on the local disk density (Kuijken \& Gilmore 1989; Bahcall, Flynn \& Gould 1992).
Therefore, non-standard halo models which allow 100 \% brown dwarf halos does not conflict with the constraints from the rotation curve or the local disk density.
This conclusion is not affected by small changes in the halo parameters.

Why does the MACHO mass become so small when the non-standard halos are considered ?
The substellar MACHO mass is obtained mainly because of the small velocity dispersion in the non-standard halos.
The timescale $\hat{t}$, which is the only observable used for the mass determination, is given by $\hat{t}=2R_{\rm E}/v_{\perp}$.
The Einstein ring diameter is proportional to $\sqrt{m}$, and hence, $m\propto v_{\perp}^2$ if the timescale is given observationally.
Consequently, the MACHO mass becomes small if the velocity dispersion in the halo is sufficiently reduced.
In fact, the tangential dispersion in the model P halo is 82 km s$^{-1}$ at $r=20$ kpc, which is less than half of the tangential dispersion for the standard halo, $\sim 200$ km s$^{-1}$.

\section{Likelihood Analysis}

In this section we perform the maximum likelihood analysis and
investigate more precisely the possible range of the MACHO mass and the halo MACHO fraction.
The method applied here is basically the same as Alcock {\it et al.}(1997), whereas the halo models are the non-standard models described in section 2.
The likelihood of finding a set of $N_{\rm obs}$ detected events with duration times $\hat{t}_{i}, i=1,...,N_{\rm obs}(=6)$, is written as
\footnote{The expression in eq.(10)
  is different from eq.(13) of Alcock {\it et al.} (1997) by the factor
  of $(\tilde{N}(m))^{N_{\rm obs}}$. Without this factor, $L$ does not follow the Poisson distribution correctly.}
\begin{equation}
  L(m,f)=(\tilde{N}(m))^{N_{\rm obs}}\exp\left[-f\tilde{N}(m)\right]
  \prod^{N_{\rm obs}}_{i=1}\left[f E\varepsilon(\hat{t}_{i})
    \frac{\displaystyle d\Gamma}{\displaystyle d\hat{t}}(\hat{t}_{i};m)
  \right],
  \label{eq:likelihood}
\end{equation}
where $\tilde{N}(m)$ is the number of events expected from Poisson statistics (see eq.[8] in Alcock {\it et al.} 1997), $f$ is the halo MACHO fraction, $E=1.82\times 10^7$ star-yr is the total exposure and $\varepsilon$ is the observational efficiency.
For the observational efficiency we use the photometric efficiency of the MACHO year-1+2 observation given in figure 8 of Alcock {\it et al.} (1997).
The likelihood contours for model H are shown in figure 3a, and those for
model P in figure 3b.  The estimated MACHO mass $m$ is
$0.06^{+0.10}_{-0.04}M_{\sun}$ for model H, and
$0.05^{+0.06}_{-0.03}M_{\sun}$ for model P (the errors denote 68\%
confidence level).  Most likely masses for both models are consistent
with brown dwarfs.  For the halo MACHO fraction $f$, the elongated contours in figure 3 make the possible range of $f$ wide, but most likely values are fairly close to unity (0.92 for model H, and 0.86 for model P).
The results for $f$ as well as the MACHO mass $m$ indicate that these halos can solely consist of brown dwarfs, confirming what we have found in the previous section.

\section{Discussion}

We have seen that a brown dwarf halo is a possible candidate for the Galactic halo when the non-standard halo models are considered.
Assuming the halo MACHO fraction of unity for the models H and P, we obtain the MACHO mass within 50 kpc of $1.3\times 10^{11}M_\odot$ and $1.0\times 10^{11}M_\odot$.
These masses are somewhat lower than the estimate of $2_{-0.7}^{+1.2}\times 10^{11} M_\odot$ by Alcock {\it et al.}(1997), but do not contradict with each other when one considers the fact that the smaller $R_0$ adopted here tends to make $\tau$ higher for a given halo model.
Since the total masses of model H and P galaxies are $2.4\times 10^{11}M_\odot$ and $1.6\times 10^{11}M_\odot$, $70$\%$\sim 80$\% of the total mass may be in the form of brown dwarfs.
To be consistent with the star counts at high Galactic latitude (e.g., Bahcall {\it et al.} 1994), such a brown dwarf halo should have a steep IMF at the low mass end (e.g., Kan-ya {\it et al.} 1996; Graff \& Freese 1996).
Unfortunately, any physical process for producing such a steep IMF is not known yet.
However, the star formation from primordial gas may be very different from the present star formation because of the different cooling processes (e.g. Palla {\it et al.} 1983, Uehara {\it et al.} 1996), and hence such an IMF may be possible if the halo brown dwarfs are different populations from disk stars.

In order to test the possibility of a brown dwarf halo, direct observations of the halo brown dwarfs are definitely important.
Unfortunately, star counts at high Galactic latitude so far made were not deep enough to detect the halo brown dwarfs.
For instance, the star counts with HST (Bahcall {\it et al.}1994), which covers a sky area of 4.4 square arcmin with the limiting I-band magnitude of $\sim 25$ mag, detected no brown dwarf candidates.
If the I-band absolute luminosity of $m\sim 0.06M_\odot$ brown dwarfs is 18 mag (e.g., Delfosse {\it et al.} 1997), there should be $\sim 10^7$ brown dwarfs brighter than 25 mag in the vicinity of the Sun (model H assumed).
The expected number in the HST filed of 4.4 square arcmin is, therefore, only 0.3, being consistent with the negative detection.
For direct detection of the halo brown dwarfs, a deeper survey with a larger area is required.
10m-class telescopes with wide field of view, such as the SUBARU telescope with Sprime-Cam, will be useful for this kind of observation.

\acknowledgments

We are grateful to Y. Sofue, J. Jugaku, K. Ohnishi, S. Inutsuka and Y. Itoh for fruitful discussions.
We also thank the referee for helpful suggestions in improving the manuscript.
One of the authors (M.H.) was financially supported by the Japan Society for the Promotion of Science.

\clearpage

\begin{figure}
\caption{Contours for the optical depth $\tau$ (thin line) and for the MACHO mass $m$ (dotted line) in the parameter space of $M$ and $a$. (a) is for Hernquist model, and (b) for Plummer model.
Contours for $\tau$ correspond to 1.0, 1.5, ..., 3.0$\times 10^{-7}$ from left, and contours for $m$ correspond to 0.04, 0.06, ..., 0.14 $M_\odot$ from left.
In the shadowed area, $\tau$ agrees with $\tau_{\rm obs}$ of $2.1_{-0.7}^{+1.1}\times 10^{-7}$, and the MACHO mass is less than 0.1$M_\odot$.}
\end{figure}

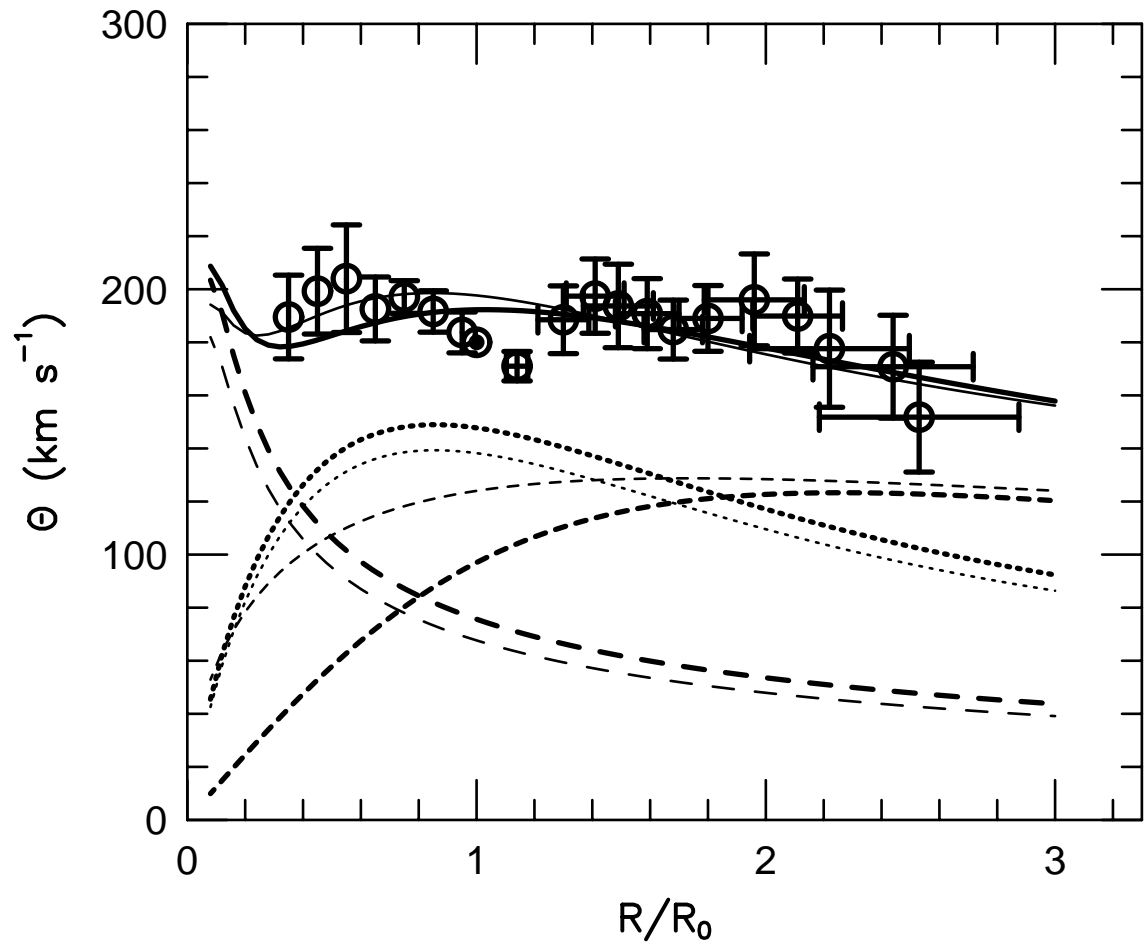
\begin{figure}
\caption{Model rotation curves for model H (thin line) and model P (thick line).
Circles with error bars show the observed rotation velocities taken from Honma \& Sofue (1997a).
The Galactic constant $\Theta_0$ is assumed to be 180 km s$^{-1}$, and the radial distance is scaled with $R_0$=7.5 kpc.
Rotation curves for each component are also shown.
Dotted lines are for disks, short dashed lines for halos, and long-dashed lines for bulges.
See text for the model parameters.}
\end{figure}

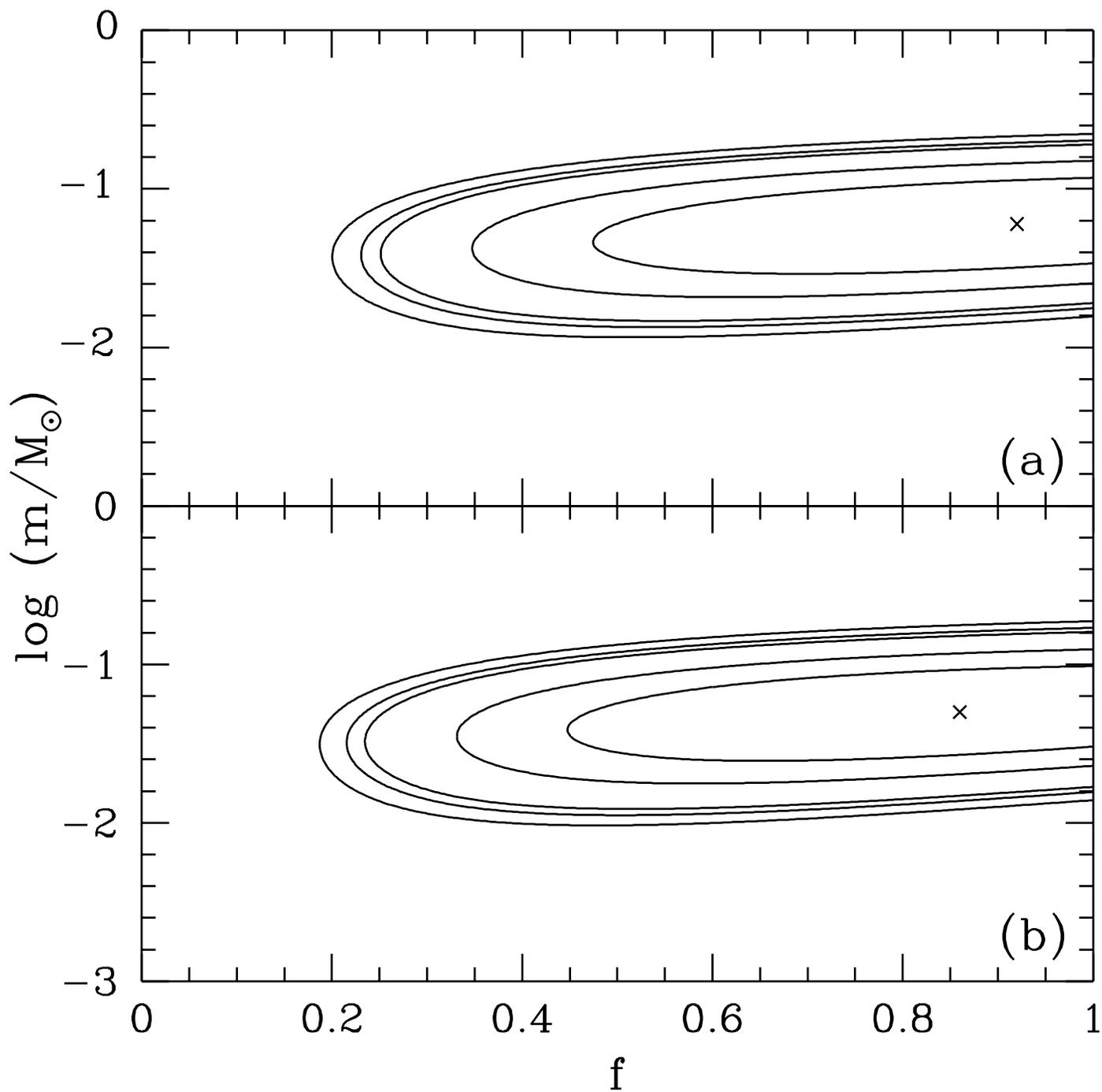
\begin{figure}
  \caption{Likelihood contours of the MACHO mass $m$ and the halo MACHO fraction $f$ for model H (a) and for model P (b), respectively. 
    The contours enclose total probabilities of 34\%, 68\%,90\%,95\%, and 99\%.
The crosses denote the peak of the likelihood.}
\end{figure}


\clearpage
\begin{references}

\reference{} Adams F. C., Laughlin G. 1996, ApJ 468, 586

\reference{} Alcock et al. 1993, Nature 365, 621

\reference{} Alcock et al. 1996, ApJ 461, 84

\reference{} Alcock et al. 1997, ApJ 486, 697

\reference{} Aubourg et al. 1993, Nature 365, 623

\reference{} Bahcall J. N., Flynn C., Gould A. 1992, ApJ 389, 234

\reference{} Bahcall J. N., Flynn C., Gould A., Kirhakos S. 1994, ApJL 435, L51

\reference{} Binney J. Tremaine S. 1987, {\it Galactic Dynamics} (Princeton University Press, Princeton).

\reference{} Bland-Hawthorn J., Freeman K. C., Quinn P. J. 1997, ApJ 490, 143

\reference{} Casertano S., van Gorkom J. H. 1991, AJ 101, 1231

\reference{} Chabrier G., Segretain L., M\'era D. 1996, ApJ 486, L21

\reference{} Charlot S., Silk J. 1995 ApJ 445, 124

\reference{} Delfosse et al. 1997, A\&A 327, L25

\reference{} Flynn C., Gould A. Bahcall J. N. 1996, ApJ 466, L55

\reference{} Gibson B. K., Mould J. R. 1997, ApJ 482, 98

\reference{} Graff D. S., Freese K. 1996, ApJ 456, L49

\reference{} Griest K. 1991, ApJ, 366, 412

\reference{} Hernquist L. 1990, ApJ 356, 359

\reference{} Honma M., Sofue Y. 1996, PASJ 48, L103

\reference{} Honma M., Sofue Y. 1997a, PASJ 49, 453

\reference{} Honma M., Sofue Y. 1997b, PASJ 49, 539

\reference{} Kan-ya Y., Nishi R., Nakamura T. 1996, PASJ 48, 479

\reference{} Kuijken K., Tremaine S. 1994, ApJ 421, 178

\reference{} Kuijken K., Gilmore G. 1989, MNRAS 239, 605

\reference{} Olling, R. P., 1996, AJ 112, 457

\reference{} Olling R. P., Merrifield M. R. 1998, MNRAS in press

\reference{} Paczynski B. 1986, ApJ 304, 1

\reference{} Palla F., Salpeter E.E., Stahlar S.W. 1983, ApJ 271, 632 

\reference{} Scalo J. M. 1986, Fund. Cosmic. Phys. 11, 1

\reference{} Uehara H., Susa H., Nishi R., Yamada M., Nakamura T. 1996, ApJ 43, L95

\end{references}
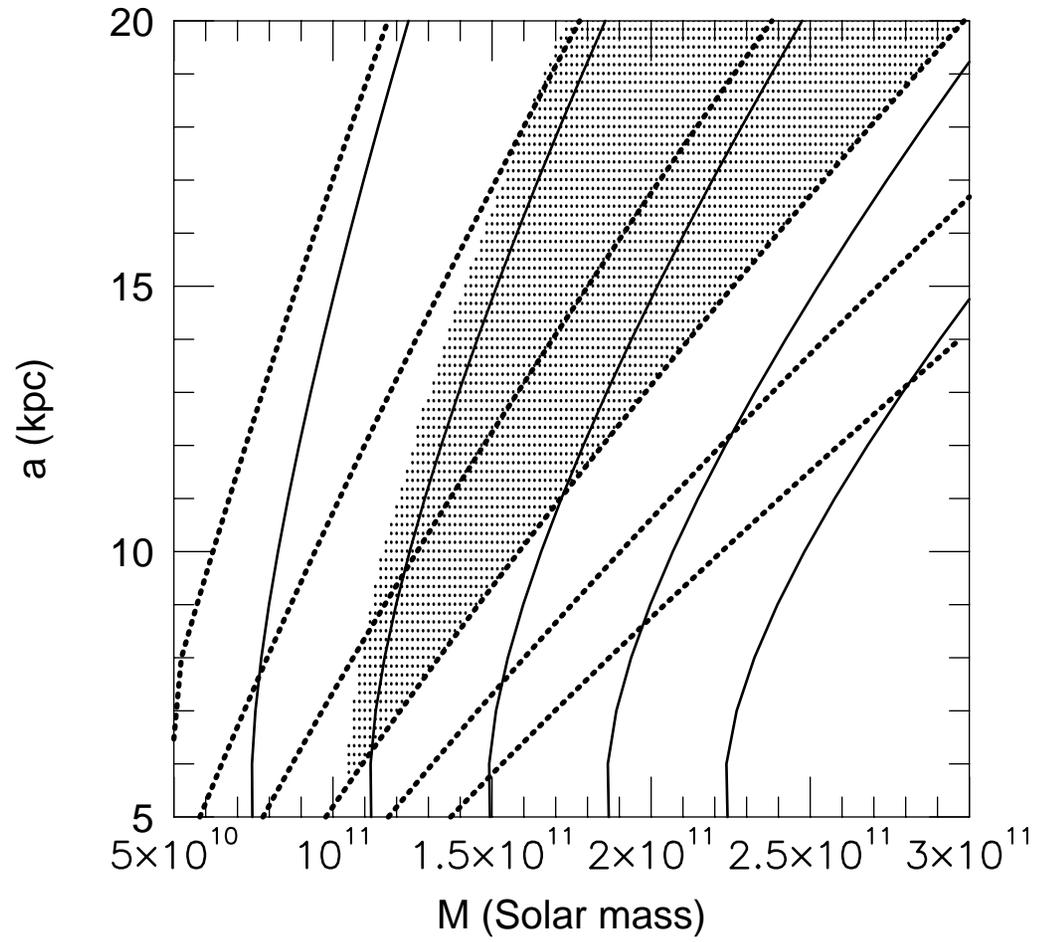
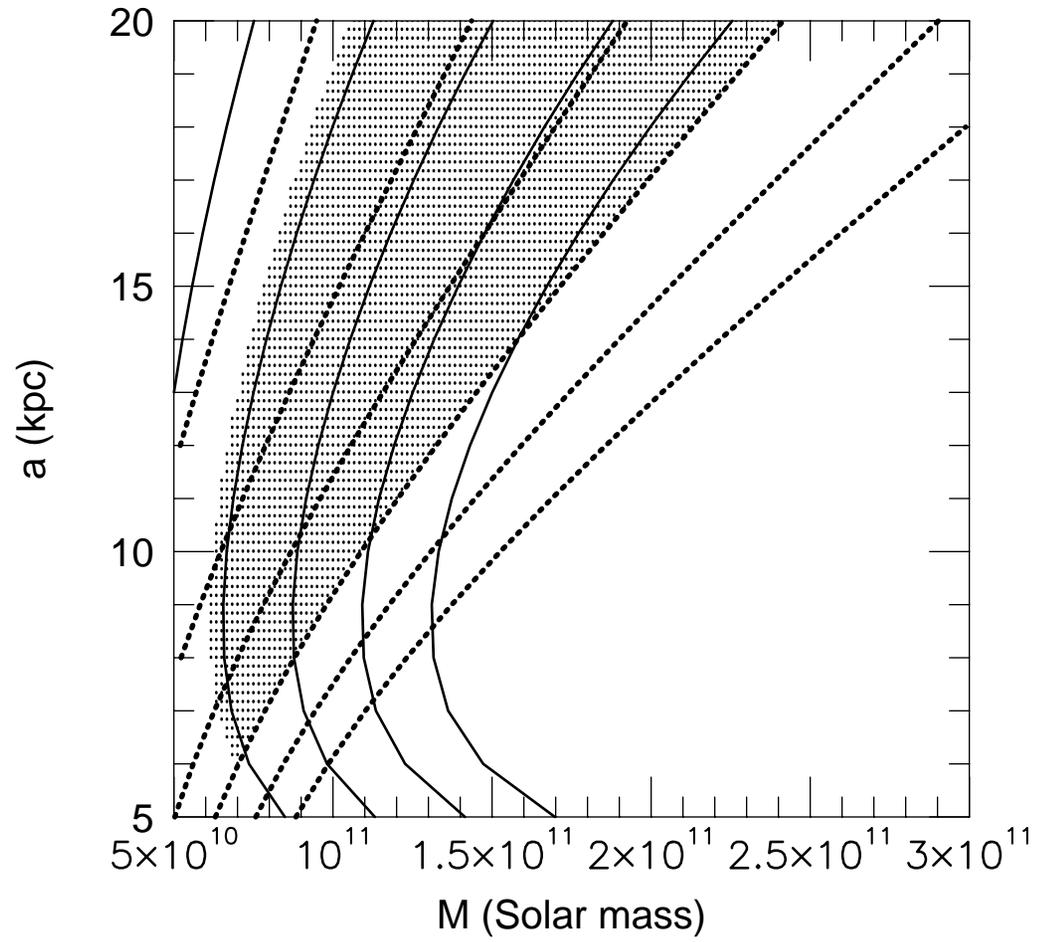
\end{document}